\documentclass[12pt]{article}
\usepackage{amsmath,graphicx}

\def\EEE{E_1{}^1}

\def\be{\begin{equation}}
\def\ee{\end{equation}}

\def\s{\text{sech}(w\tau)}
\def\t{\text{tanh}(w\tau)}
\def\sech{\text{sech}}

\def\Sm{\Sigma_-}
\def\Sc{\Sigma_\times}
\def\Nm{N_-}
\def\Nc{N_\times}
\def\Sp{\Sigma_+}
\def\SN{\mathcal{N}}

\def\dT{\partial_T}
\def\dX{\partial_X}
\def\d{\text{d}}
\def\e{\text{e}}
\def\arctanh{\text{arctanh}}
\def\Ophi{\Omega_\phi}

\def\parb{\boldsymbol{\partial}}
\def\Udot{\dot{U}}

\begin{document}

\begin{center}
{\Large\bf Spikes and matter inhomogeneities in massless scalar field models}
\vspace{.3in} \\ 
{\bf A A Coley}, 
\\Department of Mathematics \& Statistics, Dalhousie University,\\
Halifax, Nova Scotia, Canada B3H 3J5
\\Email: aac@mathstat.dal.ca
\vspace{.1in}
\\ {\bf W C Lim},
\\Department of Mathematics, University of Waikato, Private Bag 3105, Hamilton 3240, New Zealand
\\Email: wclim@waikato.ac.nz
\vspace{.1in}
\vspace{0.2in}

\end{center}

\begin{abstract}
We shall discuss the general relativistic generation of spikes in a massless scalar field or stiff perfect fluid model.
We first investigate orthogonally transitive (OT) $G_2$ stiff fluid spike models both heuristically and numerically, 
and give a new exact OT $G_2$ stiff fluid spike solution.
We then present a new two-parameter family of non-OT $G_2$ stiff fluid spike solutions, 
obtained by the generalization of non-OT $G_2$ vacuum spike solutions to the stiff fluid case
by applying Geroch's transformation on a Jacobs seed.
The dynamics of these new stiff fluid spike solutions is qualitatively different from that of the vacuum spike solutions, 
in that the matter (stiff fluid) feels the spike directly and the stiff fluid spike solution can end up with a permanent spike.
We then derive the evolution equations of non-OT $G_2$ stiff fluid models, including a second perfect fluid, in full generality, 
and briefly discuss some of their qualitative properties and their potential numerical analysis. 
Finally, we discuss how a fluid, and especially a stiff fluid or massless scalar field, affects the physics of the generation of spikes.

\end{abstract}


\section{Introduction}

It is a general feature of solutions of partial differential equations (PDE) that spikes occur~\cite{inbook:Wei2008}. 
Spikes occur in generic solutions of the Einstein field equations (EFE) of general relativity (GR)~\cite{thesis:Lim2004}.  
Indeed, when a self-similar solution of the EFE is unstable, spikes can arise near such solutions.

Spikes, originally found in the context of vacuum orthogonally transitive (OT) $G_2$ models~\cite{art:BergerMoncrief1993,art:RendallWeaver2001,art:Lim2008,art:Limetal2009}, 
describe a dynamic and spatially inhomogeneous gravitational distortion. 
Berger and Moncrief first discovered spikes in their numerical simulations~\cite{art:BergerMoncrief1993}. 
Rendall and Weaver~\cite{art:RendallWeaver2001} discovered a composition of two transformations that can map spike-free solutions to solutions with spikes. 
Using the Rendall-Weaver transformation, Lim discovered an exact OT $G_2$ spike solution~\cite{art:Lim2008}.
Recently, this solution was generalized to the non-OT $G_2$ case by applying Geroch's transformation on a Kasner seed~\cite{art:Lim2015}. 
The new solution contains two more parameters than the OT $G_2$ spike solution.

The mechanism of spike formation is simple -- the state-space orbits of nearby worldlines approach a saddle point; if
this collection of orbits straddle the stable manifold of the saddle point, then one
of the orbits becomes stuck on the stable manifold of the saddle point and heads
towards the saddle point, while the neighbouring orbits leave the saddle point.
This heuristic argument holds as long as spatial derivative terms have negligible
effect. In the case of spikes, the spatial derivative terms do have significant
effect, and the spike point that initially got stuck does leave the saddle point
eventually, and the spike that formed becomes smooth again. 

\subsubsection*{Types of spikes}

In~\cite{art:Limetal2009}, further improved numerical
evidence was presented that spikes in the Mixmaster 
regime of $G_2$ cosmologies are transient and recurring, 
supporting the conjecture that the generalized Mixmaster 
behavior is asymptotically non-local where spikes occur.
It is believed that this recurring violation of BKL locality holds in more general spacetimes.
We have previously shown explicitly that there exist ($G_2$) recurring spikes leading to
inhomogeneities and a small residual in the form of matter perturbations~\cite{art:ColeyLim2012}. 

We are also interested in incomplete spikes.
Evolving away from the initial singularity, the oscillatory regime eventually ends when $\Omega$ is no longer negligible, and
some of the spikes are in the middle of transitioning, leaving inhomogeneous imprints on the matter result.
The residuals from an incomplete spike might, in principle, be large and thus affect structure formation.
The incomplete spikes associated with Kasner saddles points occur generically in the early Universe.

Both the incomplete spikes and the recurring spikes are potentially of physical importance. 
Saddle points, related to self-similar solutions such as the Kasner solutions and FLRW models, may also occur at late times, and may also
cause spikes/tilt that might lead to further matter inhomogeneities, albeit non-generically, which
might lead to the existence of exceptional structures on large scales. 

\subsubsection*{BKL dynamics}

Belinskii, Khalatnikov and Lifshitz (BKL) \cite{art:LK63,art:BKL1970,art:BKL1982,art:BK1981} have conjectured that within GR, the approach to the
generic (past) spacelike singularity is vacuum dominated, local, and oscillatory (i.e., Mixmaster).
Studies of $G_2$ and more general cosmological models have produced numerical evidence that the BKL conjecture generally holds except possibly at isolated points 
(surfaces in the three-dimensional space) where spiky structures (``spikes") form \cite{art:Bergeretal2001,art:vEUW2002,art:Garfinkle2004num,art:Anderssonetal2005}. 
These spikes become ever narrower as the singularity is approached. 
The presence of such spikes violates the local part of the BKL conjecture.

BKL considered the EFE in synchronous coordinates and by dropping all spatial derivatives,
which geometrically corresponds to neglecting the Ricci 3-curvature of the spatial surfaces of the
synchronous coordinate system, as well as all matter terms [1]. This procedure leads to a set of
ordinary differential equations (ODEs) that are identical to those obtained in the vacuum case by
imposing spatial homogeneity and an associated simply transitive Abelian symmetry group, which
results in the vacuum Bianchi type I models whose solution is the well-known Kasner solution.
But in the general inhomogeneous context the constants of integration that appear in the Kasner
solution are replaced by spatially dependent functions, leading to a generalized Kasner solution 
(even though it is not a solution to the EFE, it is a building block when one attempts to construct generic asymptotic
solutions). 

\subsubsection*{The influence of matter}

In their seminal work, BKL \cite{art:BK1981} studied the influence of matter upon the behavior of the general inhomogeneous
solution of the EFE in the vicinity of the
initial singularity. 
In a space filled with a perfect fluid with the equation of
state $p=(\gamma-1)\rho$, for $1\leq \gamma<2$
the oscillatory regime, as the singular point
is approached asymptotically,  remains the same as in vacuum.

However, for the ``stiff matter" equation of state, $\gamma=2$, we have that $p_{\phi}=\rho_{\phi}$ and neither the
Kasner epoch nor an oscillatory regime can exist in the neighborhood of the singularity.
Indeed, it has been shown~\cite{art:BK1981} that the influence of the ``stiff matter" or a massless scalar field with $\rho_{\phi}=p_{\phi}= -g^{ab}{\phi}_{,a} {\phi}_{,b}$
results in the Jacobs relations \cite{book:Coley2003,art:CarrColey1999}:
\begin{equation}
p_{\left(  1\right)  }+p_{\left(  2\right)  }+p_{\left(  3\right)  }=1,\text{
\ \ \ \ }p_{\left(  1\right)  }^{2}+p_{\left(  2\right)  }^{2}+p_{\left(
3\right)  }^{2}=1-p^{2}\,,\label{4}
\end{equation}
where $p^{2}$ is an arbitrary time-independent function with $p^{2}<1$, for which the energy density is asymptotically of the form $\rho_{\phi} = \frac{1}{2} {\dot{\phi}^2}= p^2 t^{-2}$
(where $\phi=p\ln t$ in comoving time).
Therefore, unlike the Kasner relations, it is
possible for all three exponents $p_{\left(  a\right)  }$ to be positive simultaneously. Consequently, even if the contraction of space
starts with a quasi-Kasner epoch (\ref{4}) in which one of the exponents
$p_{\left(  a\right)  }$ is negative, the power law asymptotic behavior with
all positive exponents results after a finite number of
oscillations and then persists up to the singular point, and in general the collapse is
described by monotonic (but anisotropic) contraction along all spatial directions \cite{art:BK1981}.

\subsubsection*{Overview}

In this paper we wish to extend previous results to the massless scalar field/stiff perfect fluid case.
In the next section we briefly review massless scalar fields
and the techniques that have been used to generate exact 
stiff fluid solutions. As motivation we first
generalize the OT $G_2$ vacuum spike solution to obtain
a new exact OT $G_2$ stiff fluid spike solution, and
analyse OT $G_2$ stiff fluid spike models both heuristically and numerically~(see Section~\ref{sec:2}).

We then discuss non-OT $G_2$ stiff fluid spike solutions~(see Section~\ref{sec:3}).
We first obtain a new class of exact non-OT $G_2$ stiff fluid spike solutions. 
This is achieved, generalizing~\cite{art:Lim2015}, by 
applying the stiff fluid version of Geroch's transformation on a Jacobs seed. 
The new solution contains two more parameters than the 
OT $G_2$ stiff fluid spike solution described earlier.
We discuss these solutions.

We subsequently discuss the non-OT $G_2$ stiff fluid models in full generality~(see Section~\ref{sec:4}). 
We extend the analysis to include a second perfect fluid.
We derive the evolution equations using different normalizations and gauge choices. 
In particular, the discovery of the exact non-OT $G_2$ stiff fluid spike solution motivates the use of the fluid-comoving gauge. 
We briefly discuss some of the qualitative properties of these models (primarily to illustrate any new features of the models)
and discuss their numerical analysis.

In the final section we discuss the physical consequences of a stiff fluid or massless scalar field in the general relativistic generation of spikes.

\section{Massless scalar field}
\label{sec:2}

Scalar fields are ubiquitous in the early universe in modern theories of theoretical physics.
In the approach to the singularity it is known that the scalar field is dynamically massless \cite{book:Coley2003,art:CarrColey1999}.
Thus including massless scalar fields in early universe cosmology is important.
The field equations of a minimally coupled scalar field with timelike gradient
are formally the same as those of an irrotational stiff fluid.

We shall
concentrate on showing how spikes generate matter overdensities in a radiation fluid in general relativity in a
special class of inhomogeneous models in the initial
regime of general massless scalar field cosmological models.  
In the initial oscillatory vacuum regime, we recall that spikes recur.
We also wish to study the residual imprints of the spikes
on matter inhomogeneities in the early universe in scalar field models. As the spike inhomogeneities form,
matter undergoes gravitational instability and begins to collapse to form overdensities.

We shall {\em{normalize}}  using a $D$-normalization; when utilizing the exact solutions obtained from a 
Geroch transformation, $D$ is chosen to be the scale dependent determinant of the metric (see section 3.1).
In the OT case under consideration below, $D$-normalization is equivalent to $\beta$-normalization. Hence using
$\beta$-normalization implies that the normalized stiff fluid density $\Ophi$ ($\sim \rho_\phi \beta^{-2}$)  is a constant
(and we can then omit its trivial evolution equation).

\subsection{OT $G_2$ spike imprint analysis}

The exact OT $G_2$ stiff fluid spike solution (which can be used as the zeroth order solution in the linearization)
obtained as a simple generalization of \cite{art:Lim2008} is:
\be
\label{spike}
        (\Sm,\Nc,\Sc,\Nm) =
        \left(-c \Sm{}_\text{Taub} -\frac{1}{\sqrt{3}},
        s\Nm{}_\text{Taub},
        c \Nm{}_\text{Taub},
        -s \Sm{}_\text{Taub}
        \right).
\ee
\be
\label{csf}
        c = \frac{f^2-1}{f^2+1},\quad
        s = \frac{2f}{f^2+1},\quad
        f =w e^\tau \s x.
\ee
\be
        \Sm{}_\text{Taub}=\frac{w}{\sqrt{3}}\t-\frac{1}{\sqrt{3}},\quad
	\Nm{}_\text{Taub}=\frac{w}{\sqrt{3}}\s	
\ee
\be
	\Omega_\phi = \text{const.},\quad v_\phi = 0.
\ee
The variables there are $\beta$-normalized, where $\beta$ is the area expansion
rate of the $(y,z)$ plane and is related by $H$ = $\beta(1-\Sp)$.   
$\Sp$, $\Sm$ and $\Sc$ are components of the $\beta$-normalized rate of shear;
$\Nc$ and $\Nm$ are components of the $\beta$-normalized spatial
curvature ($\Sp$ and $q$ etc. are given in \cite{thesis:Lim2004}).
$\Omega_\phi$ is the $\beta$-normalized stiff fluid density; $v_\phi$ is the
relative stiff fluid velocity (tilt) in the $x$-direction.

More importantly, we note that (here we shall use the sign convention $\beta>0$, as opposed to~\cite{art:ColeyLim2012})
\be
        \beta = \frac12 \s e^{[\frac{w^2+7+3\Ophi}{4}]\tau - \frac14 \lambda_2} (f^2+1)^{-\frac12}.
\label{beta}
\ee

The full OT $G_2$ evolution equations which include both the stiff fluid and another perfect fluid $(\Omega,v)$ are presented in the Appendix (see Appendix D of \cite{thesis:Lim2004} 
with $A=0$), from which we can obtain the linearized evolution equations. $\gamma$ is the equation of state parameter, with $\gamma=\frac43$ describing the radiation fluid.
The evolution equations for $\beta$ and $\Omega$ are:
\begin{eqnarray}
\partial_\tau \ln\beta &=  &\frac34 [1+\Sm^2+\Sc^2+\Nc^2+\Nm^2 + (\gamma -1) \Omega + \Ophi]  \label{beta1}\\
\partial_\tau \ln\Omega & = & \frac12 \gamma v  \EEE \partial_x \ln\Omega + \frac12
\gamma \EEE \partial_x v \nonumber \\
&& - \frac34 (2 - \gamma) [1 + \Sm^2 + \Sc^2 + \Nc^2 + \Nm^2 - \Omega + \Ophi].\label{omega1}
\end{eqnarray}

In the above equations we see that the constant parts of the evolution equations for
$\beta$ and $\Omega$ are ``renormalized" by the factor $(1+\Ophi)$. Therefore, the numerics will show numerical evidence of spikes and their influence on 
matter perturbations, and the quasi-analytical results will yield similar results to those in previous papers 
on the vacuum case \cite{art:Lim2008,art:ColeyLim2012} with the renormalization of the constant
parameters.

\subsubsection*{Heuristics}

In the vacuum case ${\beta} \equiv {\beta}^\text{vac}$ and ${\Omega} \equiv {\Omega}^\text{vac}$
(given in terms of ${\beta}^\text{vac}$) are given by equations (7) and (11) in \cite{art:ColeyLim2012}, respectively.
The stiff fluid equivalent for ${\beta}$ is given by equation~(\ref{beta}), where
\be
	\beta=\beta^\text{vac}e^{\frac{3}{4}\Ophi \tau},
\ee
and the evolution equations for $\beta$, $\Omega$ are given by equations (\ref{beta1}) -- (\ref{omega1}) above.

Treating the (radiation $\gamma =  \frac{4}{3}$) $\Omega$ field as a test fluid (with negligible $\Omega$ and $v$), we obtain
\be 
	\partial_\tau \ln \Omega = \partial_\tau \ln \beta^{-(2-\gamma)}.
\ee
Again, remarkably, we can integrate this to exactly obtain
\be
	\Omega=\Omega^\text{vac}e^{-\frac{3}{4} (2 - \gamma) \Ophi \tau}.
\ee
We recall that $\tau$ increases towards the singularity, so that 
$\tau$ decreases to the future. Therefore, $\Omega$ is amplified to the future (relative to $\Omega^\text{vac}$).
We recall from~\cite{art:ColeyLim2012} that the cumulative effect of a complete spike transition
on the spatial inhomogeneity of $\beta^\text{vac}$ (and hence $\Omega^\text{vac}$) is zero.
In the stiff fluid case, $\beta$ and $\Omega$ above only differ from $\beta^\text{vac}$ and $\Omega^\text{vac}$ by a purely time-dependent factor,
so that a complete spike transition has no permanent inhomogeneous imprint on the test matter in the stiff fluid case either.

We next consider the linearized equations (which can be obtained from the Appendix), where
the zeroth order terms in the linearized equations are satisfied 
identically by the exact spike background solution. Assuming a small $\Omega_0$ (and neglecting tilt $v$), we obtain (as above)
\be
	\Omega_0=\Omega^\text{vac}e^{-\frac{3}{4} (2 - \gamma) \Ophi \tau}.
\ee
For the larger $\Omega$ case, with $\Omega_0 \neq 0$, and writing $\Omega = \Omega_0 (1 + \Omega_1)$
where $\Omega_1$ is treated as a perturbation, we obtain \cite{art:ColeyLim2012}
\be 
	\Omega_1 = \hat{\Omega}_1(x) - \frac{4}{3 \gamma(2-\gamma)} \bigg[ \int \Omega_0 d \tau  \bigg]^{-1}
\ee
or
\be 
	\Omega_1 - \hat{\Omega}_1(x) \sim \Omega^\text{vac}_1 e^{\frac{3}{4} (2 - \gamma) \Ophi \tau}.
\ee
Therefore, $\Omega_1$ is damped to the future (relative to $\Omega^\text{vac}_1$ as $\tau$ decreases). However, the overall radiation energy density is amplified to the future.

\begin{figure}
  \begin{center}
    \resizebox{\textwidth}{!}{\includegraphics{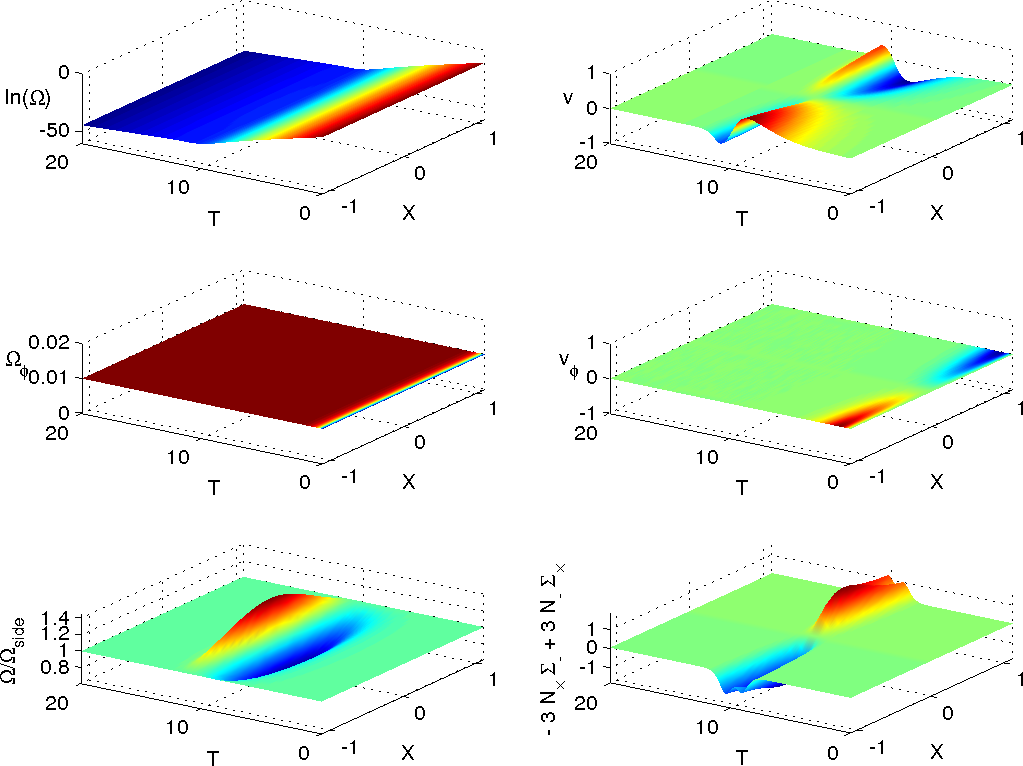}}
    \caption{Plots of $\Omega$, $v$ (for radiation fluid), $\Omega_\phi$, $v_\phi$ (for stiff fluid), the ratio $\Omega/\Omega_{X=-1.1}$ and the expression $-3\Nc \Sm + 3 \Nm \Sc$ that drives $v$.
	The plots are qualitatively the same as Figures 9 and 10 in~\cite{art:LimColey2014}, 
        showing that stiff fluid and vacuum backgrounds are qualitatively the same regarding the effect of spike on the radiation fluid.
	$\Omega_\phi$ tends to constant, while $v_\phi$ tends to zero. 
	See Equations~(\ref{IC910_1})--(\ref{IC910_4}) for the initial condition.}
    \label{fig:fig910_cropped}
\end{center}
\end{figure}

\subsubsection*{Numerics}

The system of equations in~\cite{art:LimColey2014} has been extended to include a tilted 
stiff fluid (with stiff fluid variables $\Omega_\phi$ and $v_\phi$) in the Appendix. 
It is expected that the effect of spikes on the radiation fluid to be qualitatively the same
in a vacuum or stiff fluid background.
We illustrate this by running a numerical simulation using an initial condition very similar to the one in Section 6 of~\cite{art:LimColey2014},
by specifying the initial condition $\Omega_\phi = 10^{-2}$, $v_\phi=0$.
The full initial condition (which are Equations (30)--(32) in~\cite{art:LimColey2014} but with $v=-\tanh(X/100)$) is
\begin{gather}
\label{IC910_1}
	(\Sm,\ \Nc,\ \Sc,\ \Nm) = (-c \Sm{}_\text{Taub}-\tfrac{1}{\sqrt{3}},\ s \Nm{}_\text{Taub},\ c \Nm{}_\text{Taub},\ -s \Sm{}_\text{Taub}),
\\ 
		\Omega = 10^{-5},\ v = -\tanh(X/100)),\ \Omega_\phi = 10^{-2},\ v_\phi=0, \EEE = 2,
\end{gather}
where
\be
	\Sm{}_\text{Taub} = \tfrac{1}{\sqrt{3}}[\tanh(w(T-T_0))-1],\ \Nm{}_\text{Taub} = \tfrac{w}{\sqrt{3}}\sech(w(T-T_0)),
\ee
\be
\label{IC910_4}
	c = \frac{f^2-1}{f^2+1},\ s = \frac{2f}{f^2+1},\ f =  w \sech(w(T-T_0)) (X-X_0).
\ee
We use the same parameter values as before: $w=1.5$, $T=0$, $T_0 = -10$.
The results are shown in Figure~\ref{fig:fig910_cropped}. The plots are qualitatively the same as Figures 9 and 10 in~\cite{art:LimColey2014}.
$\Omega_\phi$ quickly tends to a constant value, and $v_\phi$ quickly tends to zero.

\section{New non-OT stiff fluid spike solutions}
\label{sec:3}

We next present a new two-parameter family of non-OT $G_2$ stiff fluid spike solutions,
generalising the vacuum solutions of~\cite{art:Lim2015}. This is achieved by 
applying the stiff fluid version of Geroch's transformation~\cite{art:Geroch1971,art:Geroch1972,art:Stephani1988,art:GarfinkleGlassKrisch1997} 
on a Jacobs seed.
The new solution contains two more parameters than the OT $G_2$ stiff fluid spike solution described earlier.

Let
$g_{ab} $
be a solution of the stiff perfect fluid EFEs with energy density
$ \rho_\phi $
and pressure
$ p_\phi $ and stiff equation of state  $ p_\phi = \rho_\phi $, and fluid
four velocity $ u^a$. 
Assume that 
$ g_{ab} $
has a Killing vector field (KVF)
$ \xi ^a $.
Define the norm 
$ \lambda $
and twist
$ \omega _a $
of 
$ \xi ^a $
by
$ \lambda = {\xi ^a} {\xi _a} $
and
\be
	{\omega _a} = {\epsilon _{abcd}} {\xi ^b} {\nabla ^c} {\xi ^d}.
\ee
We assume that the KVF is orthogonal to the fluid four-velocity and 
thus
\be
	{R_{ab}} {\xi ^b} = 0.
\ee
It then follows that there is a scalar
$ \omega $
such that
$ {\omega _a} = {\nabla _a} \omega $
and that there are forms
$ \alpha _a $
and
$ \beta _a $
satisfying
\begin{gather}
 \nabla_a \omega =\varepsilon_{abcd}\xi ^b\nabla^c \xi^d,
 \\
 \nabla_{[a}\alpha_{b]} =\frac{1}{2}\varepsilon_{abcd} \nabla^c \xi^d,\quad
 \xi^a \alpha_a =\omega,
 \\
 \nabla_{[a}\beta_{b]}=2\lambda \nabla_a \xi_b + \omega \varepsilon_{abcd}  \nabla^c \xi^d,
\quad
 \xi^a \beta_a =\lambda^2 +\omega^2 -1.
\end{gather}
We solve these for $\omega$, $\alpha_a$ and $\beta_a$.
Next, we define $F$ (or $\tilde{\lambda}$) and $\eta_a$ as
\begin{align}
F = \frac{\lambda}{\tilde{\lambda}}&= (\cos\theta-\omega\sin\theta)^2 +\lambda^2 \sin^2\theta,
\\
\eta_a &=\tilde{\lambda}^{-1} \xi_a +2 \alpha_a \cos\theta\sin\theta-\beta_a \sin^2\theta,
\end{align}
for any constant $\theta$.
Then the new metric is given by
 \be
 \tilde{g}_{a b}=\frac{\lambda}{\tilde{\lambda}}(g_{a b}-\lambda^{-1} \xi_a\xi_b)+\tilde{\lambda} \eta_a \eta_b.
 \ee

This new metric is also a solution of the stiff perfect fluid EFEs
with the  same KVF. For each non-zero constant value of $\theta$ the solution is generally distinct
($\theta=0$ gives the trivial transformation $\tilde{g}_{ab} = g_{ab}$),
but it amounts to essentially adding a constant value to $\omega$. So, without loss of generality, for  $\omega \neq 0$ (and keeping an additive constant),
we can take 
$\theta = \pi/2$. 

In general, for
$ \omega \neq 0 $, the (non-tilted) stiff perfect fluid  quantities transform as follows:
\be
\tilde{\rho}_\phi = \rho_\phi / F,\quad
\tilde{u}_a = \sqrt{F} u_a,
\ee
and the determinant of the metric $g$  transforms as
\be
\tilde{g} = F^2 g.
\ee

The most relevant application of the stiff fluid version of Geroch's transformation is to generate the non-OT $G_2$ stiff fluid  spike solution.
As in~\cite{art:Lim2015}, we express a metric $g_{ab}$ using the Iwasawa frame~\cite{art:HeinzleUgglaRohr2009}, as follows. 
The metric components in terms of $b$'s and $n$'s are given by
\begin{align}
        g_{00} &= -N^2
\\
  	g_{11} &= \e^{-2b_1},\quad g_{12} = \e^{-2b_1} n_1,\quad g_{13} = \e^{-2b_1} n_2
\\
  	g_{22} &= \e^{-2b_2} + \e^{-2b_1} n_1^2,\quad g_{23} =  \e^{-2b_1} n_1 n_2 + \e^{-2b_2} n_3
\\
       	g_{33} &= \e^{-2b_3} + \e^{-2b_1} n_2^2 + \e^{-2b_2} n_3^2.
\end{align}
The seed is the Jacobs solution (stiff fluid Bianchi type I solution), 
parametrized very similarly to the vacuum case (Kasner solution) in~\cite{art:Lim2015}:
\be
   	b_1 = \frac14(w^2-1+4\rho_0)\tau,\quad
        b_2 = \frac12(w+1)\tau,\quad
        b_3 = -\frac12(w-1)\tau,\quad
        N^2 = \e^{-2b_1-2b_2-2b_3} = \e^{-\frac12(w^2+3+4\rho_0)\tau},
\ee
and $n_1=n_2=n_3=0$.
The stiff fluid density $\rho_\phi$ is simply
\be
   	\rho_\phi = \frac{\rho_0}{V^2},
\ee
where $V$ is the spatial volume, given by $V = \e^{-b_1-b_2-b_3}$.
[We note that  the stable triangular region in the Hubble-normalized $(\Sp,\Sm)$ 
plane corresponds to $0 < w < 1$, $\rho_0 > \frac14(3-w)(w+1)$.]

Following the same arguments as in~\cite{art:Lim2015}, we make a coordinate change 
and end up with a rotated Jacobs solution:
\begin{align}
\label{Jacobs_rotated}
        N^2 &= \e^{-\frac12(w^2+3+4\rho_0)\tau}
\\
  	\e^{-2b_1} &= \e^{(w-1)\tau} + n_{20}^2 \e^{-\frac12(w^2-1+4\rho_0)\tau} + n_{30}^2 \e^{-(w+1)\tau}
\\
  	\e^{-2b_2} &= \frac{\mathcal{A}^2}{\e^{-2b_1}}
\\
        \e^{-2b_3} &= \e^{-\frac12(w^2+3+4\rho_0)\tau} \mathcal{A}^{-2}
\\
  	n_1 &= \frac{n_{30} \e^{-(w-1)\tau} + n_{10} n_{20} \e^{-\frac12(w^2-1+4\rho_0)\tau}}{\e^{-2b_1}}
\\
  	n_2 &= \frac{n_{20} \e^{-\frac12(w^2-1+4\rho_0)\tau}}{\e^{-2b_1}}
\\
  	n_3 &= \e^{-\frac12(w^2-1+4\rho_0)\tau}\mathcal{A}^{-2}\left[n_{30}(n_{10}n_{30}-n_{20})\e^{-(w+1)\tau}+n_{10}\e^{(w-1)\tau} \right]
\\
\label{Jacobs_rotated_n3}
        &= \mathcal{A}^{-2}\left[n_{30}(n_{10}n_{30}-n_{20}) \e^{-\frac12[(w+1)^2+4\rho_0]\tau} + n_{10} \e^{-\frac12[(w-1)^2+4\rho_0]\tau} \right],
\intertext{where}
\label{Jacobs_area}
        \mathcal{A}^2 &= \e^{-2\tau} + n_{10}^2 \e^{-\frac12[(w-1)^2+4\rho_0]\tau} + (n_{10} n_{30} - n_{20})^2 \e^{-\frac12[(w+1)^2+4\rho_0]\tau}.
\end{align}

We now apply Geroch's transformation to the seed solution (\ref{Jacobs_rotated})--(\ref{Jacobs_rotated_n3}), using the KVF $\partial_x$.
We obtain
\be
   	\lambda = \e^{-2b_1} = \e^{(w-1)\tau} + n_{20}^2 \e^{-\frac12(w^2-1+4\rho_0)\tau} + n_{30}^2 \e^{-(w+1)\tau},\quad
        \omega = 2w n_{30} z - K y + \omega_0,
\ee
where the constant $K$ is given by
\be
   	K = \frac12 [(w-1)(w+3)+4\rho_0] n_{20} - 2 w n_{10} n_{30},
\ee
and $\omega$ is determined up to an additive constant $\omega_0$.
We could absorb $\omega_0$ by a translation in the $z$ direction if $w n_{30} \neq 0$, but we shall keep $\omega_0$ for the case $w n_{30} = 0$.
Without loss of generality, we choose $\theta=\frac{\pi}{2}$ in Geroch's transformation, so we do not need $\alpha_a$.
For $\beta_a$ we only need a particular solution.
We assume that $\beta_a$ has a zero $\tau$-component. Its other components are
\begin{align}
	\beta_1 &= \omega^2 + \lambda^2 -1
\\
  	\beta_2 &= n_{10} n_{20}^3 \e^{-(w^2-1+4\rho_0)\tau} + \left[ 2n_{10} n_{20} n_{30}^2\frac{w^2-1+4\rho_0}{(w+1)^2+4\rho_0} + 4n_{20}^2 n_{30}\frac{w+1}{(w+1)^2+4\rho_0} \right] \e^{-\frac12[(w+1)^2+4\rho_0] \tau}
\notag\\
        &\quad
              	+ 2 n_{10} n_{20}\frac{w^2-1+4\rho_0}{(w-1)^2+4\rho_0} \e^{-\frac12[(w-1)^2+4\rho_0]\tau} + (w+1) n_{30} \e^{-2\tau} + n_{30}^3 \e^{-2(w+1)\tau} + F_2(y,z)
\\
  	\beta_3 &= n_{20}^3 \e^{-(w^2-1+4\rho_0)\tau} + 2 n_{20} n_{30}^2 \frac{w^2-1+4\rho_0}{(w+1)^2+4\rho_0} \e^{-\frac12[(w+1)^2+4\rho_0] \tau}
\notag\\
        &\quad
              	+ 2 n_{20} \frac{w^2-1+4\rho_0}{(w-1)^2+4\rho_0} \e^{-\frac12[(w-1)^2+4\rho_0]\tau} + F_3(y,z)
\end{align}
where $F_2(y,z)$ and $F_3(y,z)$ satisfy the constraint equation
\be
   	- \partial_z F_2 + \partial_y F_3 + 2(w-1)\omega = 0.
\ee
For our purpose, we want $F_3$ to be as simple as possible, so we choose
\be
   	F_3 = 0,\quad F_2 = \int 2(w-1)\omega \d z = 2w(w-1) n_{30} z^2 - 2 (w-1) K y z +2(w-1)\omega_0 z.
\ee
Geroch's transformation now yields the desired metric $\tilde{g}_{ab}$, given by:
\begin{align}
\label{nonOT_spike}
        \tilde{N}^2 &= N^2 (\omega^2+\lambda^2)
\\
  	\e^{-2\tilde{b}_1} &= \frac{\e^{-2b_1}}{\omega^2+\lambda^2}
\\
  	\e^{-2\tilde{b}_2} &= \e^{-2b_2} (\omega^2+\lambda^2)
\\
\label{nonOT_spike_b3}
        \e^{-2\tilde{b}_3} &= \e^{-2b_3} (\omega^2+\lambda^2)
\\
  	\tilde{n}_1 &= -2w(w-1) n_{30} z^2 + 2 (w-1) K y z - 2(w-1)\omega_0 z
\notag\\
        &\quad
              	+ \frac{\omega^2}{\lambda}(n_{30} \e^{-(w+1)\tau} + n_{10} n_{20} \e^{-\frac12(w^2-1+4\rho_0)\tau})
\notag\\
        &\quad  -\Bigg[ n_{30} w \e^{-2\tau} + n_{10} n_{20} \frac{(w+3)(w-1)+4\rho_0}{(w-1)^2+4\rho_0} \e^{-\frac12[(w-1)^2+4\rho_0]\tau}
\notag\\
        &\quad
              	+ n_{20} n_{30} (n_{10} n_{30} - n_{20}) \frac{(w-3)(w+1)+4\rho_0}{(w+1)^2+4\rho_0} \e^{-\frac12[(w+1)^2+4\rho_0]\tau} \Bigg]
\intertext{}
  	\tilde{n}_2 &= n_{20} \Bigg[ \frac{\omega^2}{\lambda} \e^{-\frac12(w^2-1+4\rho_0)\tau} - \frac{(w+3)(w-1)+4\rho_0}{(w-1)^2+4\rho_0} \e^{-\frac12[(w-1)^2+4\rho_0]\tau}
\notag\\
        &\quad
              	- n_{30}^2 \frac{(w-3)(w+1)+4\rho_0}{(w+1)^2+4\rho_0} \e^{-\frac12[(w+1)^2+4\rho_0]\tau} \Bigg]
\\
  	\tilde{n}_3 &= \mathcal{A}^{-2} \left[ n_{10} \e^{-\frac12[(w-1)^2+4\rho_0]\tau} + n_{30} (n_{10} n_{30} - n_{20}) \e^{-\frac12[(w+1)^2+4\rho_0]\tau} \right],
\label{nonOT_spike_n3}
\end{align}
and $\mathcal{A}$, given by (\ref{Jacobs_area}), is the area density~\cite{art:vEUW2002} of the $G_2$  orbits.
The matter density for the stiff spike is
\be
	\tilde{\rho}_\phi = \frac{\rho_0}{\tilde{N}^2} = \frac{\rho_0 \e^{\frac12(w^2+3+4\rho_0)\tau} }{ (2w n_{30} z - K y + \omega_0)^2 + (\e^{(w-1)\tau} + n_{20}^2 \e^{-\frac12(w^2-1+4\rho_0)\tau} + n_{30}^2 \e^{-(w+1)\tau})^2}.
\ee

We shall focus on the case where $K=0$, or equivalently, where
\be
\label{n20choice}
        n_{20} = \frac{4w}{(w+3)(w-1)+4\rho_0} n_{10} n_{30},
\ee
which turns off the $R_2$ frame transition (which is shown to be asymptotically suppressed in~\cite{art:HeinzleUgglaRohr2009}), and eliminates the $y$-dependence.

The dynamics of the ($K=0$) stiff fluid spike solution is qualitatively different from that of the vacuum spike solution, in that the stiff fluid spike solution can end up with a permanent spike.
To produce a permanent spike, $\lambda$ must tend to zero as $\tau$ tends to infinity. This means $w$ and $\rho_0$ must satisfy
\be
   	1 - 4 \rho_0 < w^2 < 1,
\ee
assuming that $w n_{10} n_{30} \neq 0$.
The vacuum spike solution cannot meet this condition because $\rho_0=0$.

\begin{figure}
  \begin{center}
    \resizebox{\textwidth}{!}{\includegraphics{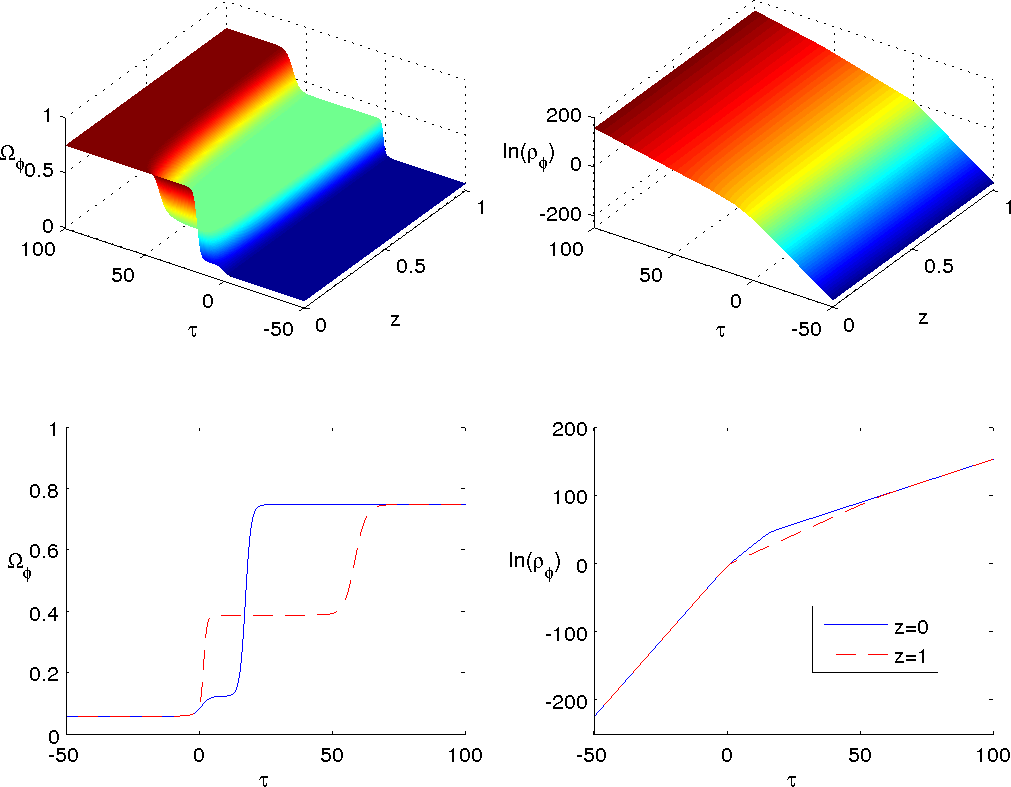}}
    \caption{Plots of Hubble-normalized $\Omega_\phi$ and $\ln(\rho_\phi)$ for the exact stiff spike solution, with $w=\frac13$, $n_{10}=0.001$, $n_{30}=1$, $\rho_0 = 0.1$. 
		In the second row of figures, solid line is for the spike worldline $z=0$, dashed line is for $z=1$.
		The plots show the spike structure in the density of the stiff fluid.}
    \label{fig:rho_stiff_cropped}
\end{center}
\end{figure}

For the exact stiff fluid spike solution, although the $\beta$-normalized $\Omega_\phi$ is independent of $z$, the Hubble-normalized $\Omega_\phi$ and physical $\rho_\phi$ do depend on $z$.
Figure~\ref{fig:rho_stiff_cropped} shows the spatial dependence of the Hubble-normalized $\Omega_\phi$ and 
$\ln(\rho_\phi)$ (plotted in coordinate $z$ without zooming). 
The inhomogeneity in $\beta$-normalized $\Omega_\phi$ is more pronounced. During the spike transition, $\rho_\phi$ is larger at the spike point.

\subsubsection*{Normalization and gauge}

We can normalize our variables with appropriate powers of a scale dependent quantity
$D$. When utilizing the exact solutions obtained from a Geroch transformation, and observing the transformation rules for the stiff fluid 
parameters above in terms of $F$, we see that an appropriate invariant choice for $D$ is the determinant of the metric
which is related to the spatial volume $V$, which then implies that the $V$-normalized stiff fluid density $\Ophi$ ($\sim \rho_\phi V^2$) is a constant
(and we can then omit its trivial evolution equation). In the OT case 
for a comoving stiff fluid seed solution (or vacuum), we have that 
${{\tilde \beta}^2} = \beta ^2 F^{-1}$, and so in this case $V$-normalization is equivalent to $\beta$-normalization.
In the non-OT case, unfortunately $V$-normalization is not equivalent to $\beta$-normalization, and $V$-normalization fails to present
self-similar solutions as equilibrium points in the $V$-normalized state space. Therefore we abandon $V$-normalization and use either $\beta$-normalization or Hubble-normalization.

We shall use the stiff fluid comoving gauge. 
In previous work (e.g., see the Appendix) we have used the area time gauge.                                                                               
Although the exact solution above satisfies both gauges, it is better to use the
stiff fluid comoving gauge because it makes matching with numerical simulations easier                                                             
i.e., it is possible to set $N = -V$ (which matches the exact solution) numerically in stiff fluid 
comoving gauge, whereas in the area time gauge, it is only possible to 
set $N=-1/\beta$, which does not match the exact solution, as $V$ is not equivalent to $\beta$ in the non-OT case.   
The evolution equations in fluid-comoving gauge (and with Hubble-normalization and choosing the lapse 
to be the volume) are given in the next section.

\section{Non-OT $G_2$ stiff fluid evolution equations}
\label{sec:4}

The discovery of the non-OT $G_2$ stiff fluid spike solution motivates the use of the fluid-comoving gauge.
For the ease of analytical analysis, it is necessary to use a normalization that presents 
solutions with timelike homothetic VF as equilibrium points.
This means using the Hubble normalization or its variation the $\beta$-normalization.
For the purpose of numerical simulation, there is no need to use such a normalization, but 
such a Hubble-normalization is convenient.
There is one requirement for numerical simulation; namely, the system should be first order.
There are two problematic terms.
The first problematic term is $\parb_3 q$ in $\parb_0 r_3$, 
which contains $\parb_3 \parb_3 \Udot_3$, a second order derivative.
We are forced to specify $r_3$ through the Codazzi constraint, so $r_3$ now contains $\parb_3 \Sp$, which is fine.
The second problematic term is $\parb_3 \Udot_3$. We cannot compute $\Udot_3$ using the 
definition of $\Udot_3$, because it would turn $\parb_3 \Udot_3$ into a second order derivative,
so we are forced to evolve $\Udot_3$. But we do not have a ready-made evolution equation for $\Udot_3$. 
We have to derive it from the definition of $\Udot_3$. This gives
\be
   	\parb_0 \Udot_3 = (q+2\Sp)\Udot_3 + 3 (\Udot_3-r_3).
\ee
Then the system is first order.

The process of normalization (while leaving the gauge unspecified) follows Section 2.3 of~\cite{thesis:Lim2004}. 
But we will just give the equations in fluid-comoving gauge (and in an Iwasawa frame).
For Hubble-normalization we can use the equations in component form given in~\cite{cal:Elst}.
The shear and $N_{\alpha\beta}$ components are:
\be
        \Sigma_{\alpha\beta} = \left( \begin{array}{ccc}
                        \Sp + \sqrt{3}\Sm & -R_3 & 0 \\
                        -R_3 & \Sp - \sqrt{3}\Sm & -R_1 \\
                        0 & - R_1 & -2\Sp
                        \end{array} \right)
        \quad
	N_{\alpha\beta} = \left( \begin{array}{ccc}
                        N_{11} & N_{12} & 0 \\
                        N_{12} & 0	& 0 \\
                        0 & 0 & 0
                        \end{array} \right)
\ee
For the sake of continuity, we shall use the ``old" variables (except $\Sigma_2$) 
where the old variables like $\Nm$, $\Nc$, $\Sc$ are related to new 
ones like $N_{11}$, $N_{12}$, $R_3$ by
\be
N_{11} = 2\sqrt{3} \Nm, N_{12} = \sqrt{3} \Nc, R_3 = - \sqrt{3} \Sc,\ R_1 = - \sqrt{3}\Sigma_2],
\ee
and the ``old" normalization (an alternative normalization 
is the new conformal normalization~\cite{art:RohrUggla2005}, which differs a bit in the $A$'s, etc.).
The evolution equations in Hubble-normalized variables are then~\cite{cal:Elst}:
\be
\label{system1_1}
        q = 2(\Sp^2+\Sm^2+\Sc^2+\tfrac13R_1^2) + 2\Omega_\phi -\tfrac13(\parb_3-r_3+\Udot_3-2A_3)\Udot_3
\ee
\be
   	N = - V,\quad   \parb_0 = -\frac{1}{VH} \partial_\tau
\ee
\be
   	\parb_3 = E_{3}{}^{3} \partial_z
\ee
\begin{align}
	\parb_0 E_{3}{}^{3} &= (q+2\Sp) E_{3}{}^{3}
\\
  	\parb_0 \Udot_3 &= (q+2\Sp)\Udot_3 + 3 (\Udot_3-r_3)
\\
  	\parb_0 A_3 &= (q+2\Sp)A_3 - (\parb_3-r_3+\Udot_3)(1+\Sp)
\\
  	\parb_0 r_3 &= (q+2\Sp)r_3 + (\parb_3-r_3+\Udot_3)(q+1)
\\
        \parb_0 \Sp &= (q-2) \Sp -2(\Nm^2+\Nc^2) +\tfrac13(\parb_3-r_3)A_3 + R_1^2 - \tfrac13(\parb_3-r_3+\Udot_3+A_3)\Udot_3
\\
        \parb_0 \Sm &= (q-2)\Sm -(\parb_3-r_3+\Udot_3-2A_3)\Nc + 2\sqrt{3} (\Sc^2-\Nm^2) - \tfrac{1}{\sqrt{3}} R_1^2
\\
        \parb_0 \Sc &= (q-2-2\sqrt{3} \Sm)\Sc + (\parb_3-r_3+\Udot_3-2A_3 - 2\sqrt{3} \Nc)\Nm
\\
        \parb_0 R_1 &= (q-2-3\Sp+\sqrt{3}\Sm)R_1
\\
        \parb_0 \Nm &= (q+2\Sp+2\sqrt{3}\Sm) \Nm + (\parb_3-r_3+\Udot_3+2\sqrt{3}\Nc) \Sc
\\
       	\parb_0 \Nc &= (q+2\Sp) \Nc -(\parb_3-r_3+\Udot_3)\Sm
\\
       	\parb_0 \Omega_\phi &= (2q-4)\Omega_\phi
\end{align}
Gauss Constraint
\be
0 = 1 + \tfrac13(2\parb_3-2r_3-3A_3)A_3 - \Nm^2 -\Nc^2 -\Sp^2-\Sm^2-\Sc^2-\tfrac13R_1^2-\Omega_\phi
\ee
Codazzi constraints
\begin{align}
        0 &= -(\parb_3-r_3)R_1 + (3A_3-\sqrt{3}\Nc)R_1
\\
        0 &= (\parb_3-r_3)(1+\Sp) -3A_3\Sp + 3\Nm\Sc - 3\Nc\Sm
\label{system1_n}
\end{align}

For this paper, we have a second tilted perfect fluid with one tilt component $(0,0,v)$. 
The equations are extended as follows:
\be
        q = 2(\Sp^2+\Sm^2+\Sc^2+\tfrac13R_1^2) + \tfrac12(\Omega+3p_\phi) + 2\Omega_\phi -\tfrac13(\parb_3-r_3+\Udot_3-2A_3)\Udot_3
\ee
\be
        G_\pm = 1 \pm (\gamma-1)v^2,\ p_\phi = \frac{(\gamma-1)(1-v^2) + \tfrac13\gamma v^2}{G_+} \Omega ,\ Q_3 = \frac{\gamma v \Omega}{G_+},\ p_\phi{}_{33} = \frac23 \frac{\gamma v^2 \Omega}{G_+}
\ee
\be
   	N = - V,\quad   \parb_0 = -\frac{1}{VH} \partial_\tau
\ee
\be
   	\parb_3 = E_{3}{}^{3} \partial_z
\ee
\begin{align}
	\parb_0 E_{3}{}^{3} &= (q+2\Sp) E_{3}{}^{3}
\\
  	\parb_0 \Udot_3 &= (q+2\Sp)\Udot_3 + 3 (\Udot_3-r_3)
\\
  	\parb_0 A_3 &= (q+2\Sp)A_3 - (\parb_3-r_3+\Udot_3)(1+\Sp)
\\
  	\parb_0 r_3 &= (q+2\Sp)r_3 + (\parb_3-r_3+\Udot_3)(q+1)
\\
        \parb_0 \Sp &= (q-2) \Sp -2(\Nm^2+\Nc^2) +\tfrac13(\parb_3-r_3)A_3 + R_1^2 - \tfrac13(\parb_3-r_3+\Udot_3+A_3)\Udot_3 - \frac{\gamma v^2 \Omega}{G_+}
\\
        \parb_0 \Sm &= (q-2)\Sm -(\parb_3-r_3+\Udot_3-2A_3)\Nc + 2\sqrt{3} (\Sc^2-\Nm^2) - \tfrac{1}{\sqrt{3}} R_1^2
\\
        \parb_0 \Sc &= (q-2-2\sqrt{3} \Sm)\Sc + (\parb_3-r_3+\Udot_3-2A_3 - 2\sqrt{3} \Nc)\Nm
\\
        \parb_0 R_1 &= (q-2-3\Sp+\sqrt{3}\Sm)R_1
\\
        \parb_0 \Nm &= (q+2\Sp+2\sqrt{3}\Sm) \Nm + (\parb_3-r_3+\Udot_3+2\sqrt{3}\Nc) \Sc
\\
        \parb_0 \Nc &= (q+2\Sp) \Nc -(\parb_3-r_3+\Udot_3)\Sm
\\
  	\parb_0 \Omega_\phi &= (2q-4)\Omega_\phi
\end{align}
Gauss Constraint
\be
0 = 1 + \tfrac13(2\parb_3-2r_3-3A_3)A_3 - \Nm^2 -\Nc^2 -\Sp^2-\Sm^2-\Sc^2-\tfrac13R_1^2-\Omega-\Omega_\phi
\ee
Codazzi constraints
\begin{align}
        0 &= -(\parb_3-r_3)R_1 + (3A_3-\sqrt{3}\Nc)R_1
\\
        0 &= (\parb_3-r_3)(1+\Sp) -3A_3\Sp + 3\Nm\Sc - 3\Nc\Sm - \frac32 \frac{\gamma v \Omega}{G_+}
\end{align}
Evolution equations for perfect fluid variables $\Omega$ and $v$:
\begin{align}
        \parb_0 \ln\Omega &= -\frac{\gamma v}{G_+} \parb_3 \ln\Omega - \frac{\gamma G_-}{G_+^2} \parb_3 v
\notag\\
                        &\quad + G_+^{-1} \left[ 2G_+ q -(3\gamma-2) -(2-\gamma)v^2 + 2\gamma v^2 \Sp - 2\gamma v(-r_3+\Udot_3 -A_3)\right]
\\
        \parb_0 v &= -\frac{(\gamma-1)}{\gamma} \frac{(1-v^2)^2}{G_-} \parb_3 \ln \Omega + \frac{[(3\gamma-4)-(\gamma-1)(4-\gamma)v^2]v}{G_+ G_-} \parb_3 v
\notag\\
                &\quad + \frac{(1-v^2)}{G_-}\left[2\frac{\gamma-1}{\gamma}(1-v^2)r_3 + (3\gamma-4)v + 2v\Sp - {G_-}\Udot_3-2(\gamma-1)v^2A_3\right]
\end{align}

\subsubsection*{Numerics}

As a first step towards a numerical analysis, we shall simulate the system~(\ref{system1_1})--(\ref{system1_n}), which is without the second perfect fluid.
For the zooming method, we shall use the dynamic zooming introduced in~\cite{art:Clarksonetal2013},
in which the outer boundary travels inward at the speed of light.
As a result the operators $\parb_0$ and $\parb_3$ are replaced by
\be
	\parb_0 = \SN^{-1}\left(\partial_T - \frac{\partial_T z}{\partial_Z z} \partial_Z\right),\quad
	\parb_3 = E_3{}^3 \frac{1}{\partial_Z z} \partial_Z,
\ee
where
\be
	T = \tau,\quad Z=Z(\tau,z)
\ee
are the new coordinates in the zoomed view, with the evolution of unzoomed coordinate $z(T,Z)$ defined by
\be
	\partial_T z = (\SN E_3{}^3)_\text{ob} \frac{Z}{Z_\text{ob}},
\ee
where the subscript $\text{ob}$ denotes evaluation at the outer boundary.
The dynamic zooming is more economical than the specified zooming of~\cite{art:Limetal2009} in that the outer boundary travels inward at exactly the speed of light.
This is important in saving numerical resources, as the horizon size grows during an $R_1$ transition.
Due to this growth, spikes appearing after the $R_1$ transition appear wider than the those before.
In order to capture the wider spike, a larger domain of simulation is needed, but the spike appearing before the $R_1$ transition would appear narrower relative to this larger domain,
and therefore more grid points are needed to provide sufficient numerical resolution for the narrower spike.

In order to ensure that the code is correct, we shall simulate the exact non-OT $G_2$ stiff fluid spike solution.
We choose the following values for the parameters:
\be
\label{nonOT_param}
	w = \frac13,\quad
	n_{10} = 0.001,\quad
	n_{30} = 1,\quad
	K = 0,\quad
	\rho_0 = 0.00075,\quad
	\omega_0 = 0.
\ee
We use the domain $Z \in [0,1]$ with 1001 uniformly spaced grid points, fine-tune the initial domain size $z$ to 6 digits by trial and error
(as the horizon size will shrink to about $10^{-6}$ of its initial size by $T=25$)
\be
	       z = 148.417 Z,
\ee
and run the simulation from $T=-5$ to $T=25$. Fine-tuning is needed because, if $z$ is too big then one loses numerical
resolution, and if $z$ is too small then the outer boundary would hit the inner boundary $z=0$ before $T=25$. 
We also limit $T$ to reduce the number of digits in the fine-tuning.

\begin{figure}
  \begin{center}
    \resizebox{\textwidth}{!}{\includegraphics{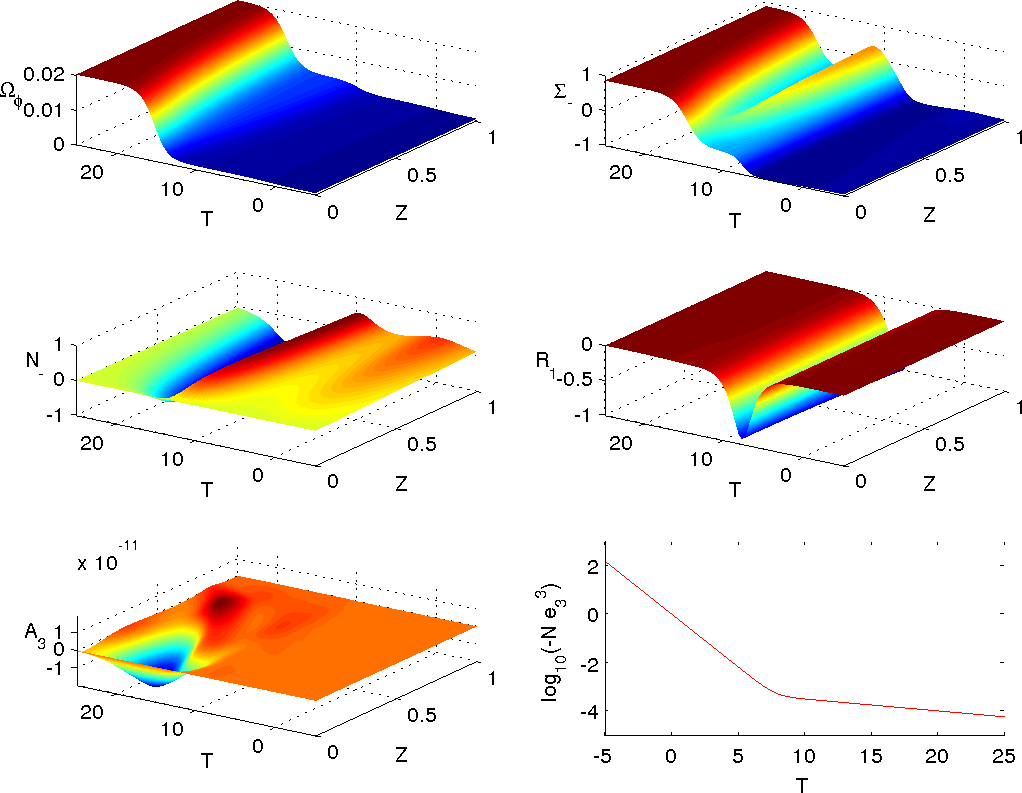}}
    \caption{Plots of $\Omega_\phi$, $\Sm$, $\Nm$, $R_1$, $A_3$ and $\log_{10}(-\SN E_3{}^3)$ for the numerical simulation of the exact solution with parameters~(\ref{nonOT_param}).}
    \label{fig:nonOT_fig_id7_cropped}
\end{center}
\end{figure}

We plot the numerical results for $\Omega_\phi$, $\Sm$, $\Nm$, $R_1$, $A_3$ and $\log_{10}(-\SN E_3{}^3)$ in Figure~\ref{fig:nonOT_fig_id7_cropped}.
It can be seen that the spike first forms at about $T=0$ and then becomes wider (in this zoomed view, but becomes narrower in the unzoomed view).
The $R_1$ transition occurs during $T\in[5,10]$, and the spike becomes narrower again, 
until it undergoes a transition during $T\in[10,20]$ and resolves.
Figure~\ref{fig:rho_stiff_cropped}, which shows the unzoomed view of the exact non-OT stiff fluid spike solution, 
hides the finer details of the spike profile and the width of the spike at different times.
Here we shall comment on the characteristic width of the horizon before and after the $R_1$ transition.
Before the $R_1$ transition, the spike solution is close to the OT solution with $w=\frac13$. Recall from~\cite{art:Lim2008} that the horizon size
decays at the rate $\e^{-\tau}$. After the $R_1$ transition, the horizon size decays at the 
rate $\e^{-0.1\tau}$ (see the last panel of Figure~\ref{fig:nonOT_fig_id7_cropped}).
Over the duration of the simulation, the horizon shrinks by a factor of $10^6$, hence the initial $z$ needs to be fine-tune to 6 digits accuracy.
We shall leave further numerical analysis to the next paper.

\section{Conclusion}

In this paper we have discussed massless scalar 
field/stiff perfect fluids in the general
relativistic generation of spikes.
We first studied OT $G_2$  stiff fluid spike models 
both heuristically and numerically, and 
generalized the vacuum solution to obtain  
a new exact OT $G_2$ stiff fluid spike solution.

We then discussed non-OT $G_2$ stiff fluid spike models.
We presented a new two-parameter family
of non-OT $G_2$ stiff fluid spike solutions, obtained by the generalization of non-OT $G_2$ vacuum spike solutions \cite{art:Lim2015} to the stiff fluid case
and achieved by applying the stiff fluid version of Geroch's transformation on a Jacobs seed.  
The dynamics of the new ($K=0$) non-OT $G_2$  stiff fluid spike solutions  is qualitatively
different from that of the vacuum spike solution, in that the matter (stiff fluid) feels the
spike directly and the stiff fluid
spike solutions can end up with a permanent spike.

We next derived the evolution equations of non-OT $G_2$ stiff fluid models, including a second perfect fluid, in full generality.
We discussed the evolution equations using different normalizations and gauge choices, 
motivated by the discovery of the exact non-OT stiff spike solutions and in order to be consistent with previous analyses. 
We shall return to the issue of different normalization and gauge choices in future work. 
We also briefly discussed some of the qualitative properties of these models and their numerical analysis
(and, in particular, the potential problems with numerical resolution).
Qualitatively we found that the spike imprint in a stiff fluid background is the same as the previous vacuum case.
We shall present a full numerical analysis (of the general equations) in a follow up paper.

We have been particularly interested in how a fluid, and especially a stiff fluid or massless scalar field, 
affects the physical consequences of the general relativistic generation of spikes. Let us briefly discuss this further.

\subsubsection*{Discussion}

In previous work \cite{art:ColeyLim2012} we explicitly showed that spikes naturally occur in a class of non-vacuum $G_2$ models and, due to
gravitational instability, leave small residual imprints on matter in the form of matter perturbations.
We have been particularly interested in recurring and complete spikes formed in the oscillatory regime (or recurring spikes for short)~\cite{art:Lim2008,art:Limetal2009}
and incomplete spikes, and their imprint on matter and structure formation.
We have obtained further numerical evidence for the existence of spikes and general relativistic matter perturbations~\cite{art:LimColey2014}, 
which support the results of \cite{art:ColeyLim2012}.
We have generalized these results to massless scalar field/stiff perfect fluid models in this paper to further illustrate the possible existence of a
general relativistic mechanism for generating matter perturbations of physical interest.

With a tilted fluid, the tilt provides another mechanism in generating matter inhomogeneities  due to the non-negligible divergence term caused by the instability in the tilt.
In~\cite{art:LimColey2014} we investigated the evolution equations of the OT $G_2$ models with a perfect fluid and
we concluded that it is the tilt instability that plays the primary role in the formation of matter inhomogeneities in these models
(while the spike mechanism plays a secondary role in generating matter inhomogeneities).

In the early Universe we have explicitly shown that there exist $G_2$ recurring spikes that lead to
inhomogeneities and a residual in the form of matter perturbations, that these occur
naturally within generic cosmology models within GR, and that they are not local but form
on surfaces and give rise to a distribution of perturbations.  
In $G_2$ models these spike surfaces are parallel and do not intersect.
In general spacetimes however, two spikes surfaces may intersect along a curve,
and this curve may intersect with a third spike surface at a point, leading to matter inhomogeneities forming on a web of surfaces, curves and points.
There are tantalising hints that filamentary structures and voids would occur naturally in this scenario.

Inflationary cosmology provides a causal mechanism which 
generates the primordial perturbations which were later responsible for the formation of 
large scale structures of our Universe due to gravitational collapse.
The density perturbations produced during inflation are due to
quantum fluctuations in the matter and gravitational fields \cite{art:Mukhanovetal1992,book:LythLiddle2009}. 
Primeval fluctuations,
which are subsequently amplified
outside the Hubble radius, are then thought to be present at the end of the inflationary
epoch. Previously, we have speculated  \cite{art:ColeyLim2012} whether 
recurring spikes might be an alternative to the inflationary mechanism for
generating matter perturbations in the early Universe.
Indeed, there are some
similarities with the perturbations and structure formation created in 
cosmic string models~\cite{art:LimColey2014};
the inhomogeneities occur on closed circles or infinite lines \cite{art:ColeyLim2012}, 
similar to what happens in the case of
topological defects.

Saddles, related to Kasner solutions and FLRW models, may also occur at late times, and may also
cause spikes/tilt that might lead to further matter inhomogeneities, albeit non-generically.
Permanent spikes
in LTB models were studied in \cite{art:ColeyLim2014}, which might offer an
alternative general relativistic spike mechanism for naturally generating (a small number of)
exceptional structures at late times.

\enlargethispage{5cm}

\section{Appendix: OT $G_2$ Equations}

The evolution equations for the OT $G_2$ model with KVFs acting on a plane, 
and with two perfect fluids (one of them stiff).
(where the coordinate variable $T$ increases towards the singularity) are \cite{thesis:Lim2004}:

\begin{align}
        \dT\ln\beta &= - AX \dX\ln\beta + \frac32(1-\Sp) - \frac34(2-\gamma)\frac{1-v^2}{G_+}\Omega
\\
        \dT\ln\EEE &= - AX \dX\ln\EEE - 1 + \frac34(2-\gamma)\frac{1-v^2}{G_+}\Omega
\\
        \dT\Sm &= - AX \dX\Sm
                + \tfrac12\e^{AT}\EEE \dX\Nc
                + \frac34(2-\gamma)\frac{1-v^2}{G_+}\Omega \Sm - \sqrt{3}(\Sc^2-\Nm^2)
\\
        \dT\Nc &= - AX \dX\Nc
                + \tfrac12\e^{AT}\EEE \dX\Sm
                - \Nc
		+ \frac34(2-\gamma)\frac{1-v^2}{G_+}\Omega \Nc
\\
        \dT\Sc &= - AX \dX\Sc
                - \tfrac12\e^{AT}\EEE \dX\Nm
		+ \frac34(2-\gamma)\frac{1-v^2}{G_+}\Omega \Sc
                + \sqrt{3}\Sm\Sc + \sqrt{3}\Nc\Nm
\\ 
        \dT\Nm &= - AX \dX\Nm
                - \tfrac12\e^{AT}\EEE \dX\Sc
                - \Nm
		+ \frac34(2-\gamma)\frac{1-v^2}{G_+}\Omega \Nm
                - \sqrt{3}\Sm\Nm - \sqrt{3}\Sc\Nc
\\
	\dT\ln\Omega &= - AX \dX\ln\Omega 
		- \frac{\gamma v}{2G_+} \e^{AT}\EEE \dX\ln\Omega
		+ \frac{\gamma G_-(1-v^2)}{2G_+^2} \dX \arctanh v
\notag\\
		&\quad- \frac{\gamma}{G_+} \left[ \frac{G_+}{\gamma}(q+1)-\frac12(1-3\Sp)(1+v^2)-1 \right]
\\
	\dT\arctanh v &= - AX \dX\arctanh v
		+ \frac{(\gamma-1)(1-v^2)}{2 \gamma G_-} \e^{AT}\EEE \dX\ln\Omega
\notag\\        
		&\quad- [3\gamma-4-(\gamma-1)(4-\gamma)v^2] \frac{v}{2 G_+ G_-} \e^{AT}\EEE \dX\arctanh v
\notag\\
		&\quad+ \frac{1}{2\gamma G_-} \left[ (2-\gamma)G_+ r - \gamma v(3\gamma-4 + 3(2-\gamma)\Sp) \right]
\\
        \dT\ln\Omega_\phi &= - AX \dX\ln\Omega_\phi
                - \frac{v_\phi}{1+v_\phi^2} \e^{AT}\EEE \dX\ln\Omega_\phi
                + \frac{(1-v_\phi^2)^2}{(1+v_\phi^2)^2} \dX \arctanh v_\phi
\notag\\
                &\quad- \frac{2}{1+v_\phi^2} \left[ \frac{1+v_\phi^2}{2}(q+1)-\frac12(1-3\Sp)(1+v_\phi^2)-1 \right]
\\
        \dT\arctanh v_\phi &= - AX \dX\arctanh v_\phi
                + \frac{1}{4} \e^{AT}\EEE \dX\ln\Omega_\phi
\notag\\
                &\quad- \frac{v_\phi}{1+v_\phi^2} \e^{AT}\EEE \dX\arctanh v_\phi
           	- \frac{v_\phi}{1-v_\phi^2}
\end{align}  
where $\Sp$, $q$, $r$, $G_\pm$ are given by
\begin{align}
	\Sp &= \frac12\left( 1-\Sm^2-\Sc^2-\Nm^2-\Nc^2-\Omega-\Ophi\right)
\\
	q &= 2-3\Sp - \frac32(2-\gamma)\frac{1-v^2}{G_+}\Omega
\\
	r &= -3\Nc\Sm + 3\Nm\Sc - \frac{3\gamma v}{2 G_+} \Omega - \frac{3 v_\phi}{1+v_\phi^2} \Omega_\phi
\\
	G_\pm &= 1 \pm (\gamma-1) v^2.
\end{align}

\section*{Acknowledgments}

This work was supported, in part, by NSERC of Canada.

\end{document}